# Minimizing hysteresis in martensite phase transforming magnetocaloric Heusler alloys

Luana Caron[1*], Parul Devi[1], Alexandre M. G. Carvalho[2], Claudia Felser[1], Sanjay Singh[1,3†]

[1]Max Planck Institute for Chemical Physics of Solids, Nöthnitzer Str. 40, 01187 Dresden, Germany

[2]Laboratório Nacional de Luz Síncrotron (LNLS), Centro Nacional de Pesquisa em Energia e Materiais (CNPEM), CEP 13083-970, Campinas, São Paulo, Brazil

[3]School of Materials Science and Technology, Indian Institute of Technology (Banaras Hindu University), Varanasi-221005, India.

The large magnetocaloric effect in Heusler alloys showing martensite phase transformation puts them forward as efficient materials for magnetic refrigeration. However, irreversibility of the magnetocaloric cooling cycle is a major challenge for real applications. This irreversibility is directly linked to the thermal hysteresis at the first-order martensite phase transition. Therefore, minimizing the hysteresis is essential in order to achieve reversibility. Here we show a large reduction in the thermal hysteresis at the martensite transition in the $Ni_2Mn_{1.4}In_{0.6}$ and $Ni_{1.8}Co_{0.2}Mn_{1.4}In_{0.6}$ Heusler alloys upon the application of hydrostatic pressure. Our pressure dependent X-ray diffraction study on $Ni_2Mn_{1.4}In_{0.6}$ reveals that with increasing pressure the lattice parameters of the two crystallographic phases (austenite and martensite) change in such a way that they increasingly satisfy the geometric compatibility (co-factor) condition. These results provide an opportunity to overcome the hysteresis problem and hence the irreversible behavior in Heusler materials using pressure as an external parameter.

Materials presenting large magnetocaloric effect (MCE) have been intensively studied aiming at applications for magnetic cooling[1-6]. Among MCE materials, shape memory Heusler alloys (SMHAs) are of great interest as their transition temperatures can be easily tuned and they do not contain rare-earth elements[2-4]. The large MCE in SMHAs is basically due to the first order austenite to martensite phase transition, which is also responsible for the shape memory phenomenon [7-11]. At the martensite phase transition these alloys undergo a change from the high symmetry austenite phase to the lower symmetry martensite phase where a large magnetization change occurs giving rise to a high MCE. However, the crystallographic change, which generates large and useful MCE's, also makes the transition less reversible. Just as to bring water from liquid to gas state one needs to lend the molecules enough energy in the form of heat, all first order phase transitions require an energy input to be driven. When this energy input is larger than the effect's yield, a cycle relying on this phenomenon will be highly inefficient if not completely irreversible, making applications unfeasible[5,12]. In the case of the martensitic phase transition in Heusler alloys this is the energy the system requires to go between the high symmetry austenite phase and the lower symmetry martensite phase. This energy input, or energy barrier, is manifested in the latent heat and thermal/magnetic hysteresis of the transition[13]. The larger these quantities, the less reversible a first order phase transition is. Therefore, it is not surprising that the current research in MCE focuses, to a great extent, on minimising thermal hysteresis as a means to improve reversibility and thus efficiency in prospective applications[14-16]. In this context, a set of rules for thermal hysteresis minimisation at the austenite-martensite phase transition in non-magnetic shape memory alloys has been developed[13,15-19]. It has been reported that the reversibility of the austenite to martensite phase transition depends basically on the compatibility between the high and low symmetry phases on either side of the structural transformation. The structural transformation taking place at the martensitic phase transition is described by the transformation stretch tensor $\mathbf{U}$, whose elements are derived from the lattice parameters of the austenite and martensite phases. The compatibility condition itself is that $\lambda_2 = 1$, where $\lambda_1 \leq \lambda_2 \leq \lambda_3$ are the ordered eigenvalues of $\mathbf{U}$. Therefore, by satisfying the $\lambda_2 = 1$ condition, thermal hysteresis and thus the energy barrier at the magneto-structural martensitic transition are reduced. Since both shape memory and the large MCE in Heusler alloys have a common origin, by achieving shape memory the reversibility of the MCE is improved.

James and co-workers[16,18] propose a composition-dependent approach to obtain different lattice parameters on the phases (and different compatibilities between them) and thus pinpoint compositions that should present low hysteresis for non-magnetic shape memory alloys. A similar study has been recently reported for magnetic Heusler alloys by Stern-Taulats and co-workers[14]. At first glance this approach is elegant in its simplicity. However, changing composition alters much more than lattice parameters, and this approach does not take into account non-intended effects such as change in electron count and structural disorder, which deeply influence the magneto-structural properties of Heusler alloys.

In this work, we study the effect of hydrostatic pressure on the thermal hysteresis on $Ni_2Mn_{1.4}In_{0.6}$ phase transforming magnetocaloric Heusler alloys. Our results show that pressure can reduce the thermal hysteresis across the martensite phase transition by approaching the compatibility condition ($\lambda_2 = 1$). Pressure is a clean mechanism as it keeps the sample composition intact, changing solely its structure and therefore the compatibility of the martensite and austenite phases. $Ni_2Mn_{1.4}In_{0.6}$ shows a phase transition between ferromagnetic (FM) cubic austenite and antiferromagnetic (AFM) 3M modulated monoclinic martensite phase just below room temperature and a Curie temperature $T_C$ at around 315 K. The change in hysteresis width of the martensite transition due to pressure was monitored through magnetization measurements and found to decrease with increasing pressure. This decrease in hysteresis with increasing pressure is explored and explained using pressure dependent X-rays diffraction (XRD), which reveals the enhancement of the compatibility condition ($\lambda_2$ approaching

1) and therefore a lower energy barrier. We also show that the same behavior is found in another important Heusler composition $Ni_{1.8}Co_{0.2}Mn_{1.4}In_{0.6}$.

The details of sample preparation, magnetization, diffraction (ambient and under pressure), structure refinement and calculation of the compatibility factor $\lambda_2$ are provided in the Supplementary Material. In the $Ni_2Mn_{1.4}In_{0.6}$ alloy at ambient pressure, the transition from the austenite to the martensite phase during cooling occurs at $T_M = 272$ K while the reverse transition, from martensite to austenite, occurs at $T_A = 283$ K due to a thermal hysteresis of approximately 11 K (see Fig. 1). Pressure stabilizes the AFM martensite phase, and $T_M$ shifts to higher temperatures at a rate of 2.8 K/kbar. This value is in good agreement with results reported on similar compositions[14,20]. However, the transition from martensite to austenite $T_A$ is less sensitive to pressure and shifts at a rate of 2.36 K/kbar, resulting in a reduction of the thermal hysteresis with increasing pressure (see inset of Fig. 1). The hysteresis is found to decrease linearly to about 60% (7.3 K for P = 9 kbar) of its original value (11 K at P = 0 kbar) at a rate of 0.46 K/kbar. $T_C$ is also found to shift to higher temperatures with increasing pressure, but at about a tenth of the rate (0.24 K/kbar) of the martensite transition, in excellent agreement with previously reported values[21]. If the trends for the shift of the critical temperatures for both the martensite and FM to PM austenite transitions with increasing pressure remain the same above 10 kbar, we estimate that the two transitions should merge at around 15 kbar for this compound, far below the pressure were the hysteresis should vanish at approximately 24 kbar.

The decrease in thermal hysteresis observed in the magnetization measurements suggests that the austenite-martensite phase compatibility is enhanced under pressure. This compatibility is quantified by the middle eigenvalue $\lambda_2$ of the transformation tensor $\mathbf{U}$, which is obtained from the lattice parameters of both phases (see the Supplementary Material for a detailed description of the matrix and its elements). Thus, to study the change of the phases compatibility under pressure we performed temperature dependent XRD under hydrostatic pressure. The lattice parameters and volume of both phases are presented in Fig. 2 at 320 K where the material is completely in the austenite phase, and at 240 K where only the martensite phase is observed. The lattice parameters and volume of both phases are found to decrease linearly with increasing applied pressure. However, the behavior of the monoclinic angle $\beta$ is found to be non-linear upon the increase of the applied pressure.

The $\lambda_2$ eigenvalue of the transformation matrix at different applied pressures was calculated from the lattice parameters shown in Fig. 2 using the transformation tensor **U** (see Supplementary Material). Interestingly, the value of $\lambda_2$ decreases with a similar trend as the thermal hysteresis (inset of Fig.1), and approaches values increasingly closer to 1 with increasing pressure (see Fig. 3a). This shows that the enhanced compatibility between the austenite and martensite phases is responsible for the decrease in thermal hysteresis with pressure. Moreover, the effect of pressure is also reflected on the latent heat of the transition and not only on the thermal hysteresis since it affects the energy barrier itself, as previously observed in a composition-tuned $\lambda_2$ study by Zhang et al[13]. Using the Clausius-Clapeyron relation the entropy change due to the structural martensitic transition can be calculated and thus the latent heat involved in the process (see the Supplementary Material for the actual derivation). The latent heat due to the structural transition (in the absence of an applied magnetic field) is found to decrease with increasing applied pressure (see Fig. 3b). Notice that, since the pressure sensitivities are different for the cooling and heating transitions, as are the transition temperatures, two sets of latent heat values are obtained. Therefore, both thermal hysteresis and latent heat decrease with increasing pressure, indicating that the energy barrier itself is decreased.

The enhanced compatibility between the phases under hydrostatic pressure can also be understood from a structural point of view by looking at the compressibility of the individual phases. The isothermal compressibility ($\beta$) of the austenite and martensite phases are $\beta_{aus} = 1.003 \times 10^{-3}$kbar$^{-1}$ and $\beta_{mart} = 0.957 \times 10^{-3}$kbar$^{-1}$, respectively, calculated from the data in the lower panel of Fig. 2. Therefore, the austenite phase is slightly more compressible than the martensite phase, which makes the lattice parameter mismatch smaller and the phases more compatible with increasing pressure, bringing $\lambda_2$ closer to unity.

Minimizing hysteresis is essential in order to achieve shape memory and, consequently, a reversible magnetocaloric effect. The lower the hysteresis and the latent heat at the phase transition the lower its energy cost is, making the magnetocaloric effect more reversible and prospective applications more efficient. For example, in the case of magnetocaloric-based refrigeration, the amount of heat that can be extracted, also called refrigeration capacity (RC), is given by the area below the entropy change vs. temperature curve. However, when using a material presenting a first order phase transition in a refrigeration cycle, this quantity corresponds to the area of the overlap between the entropy change vs. temperature curves measured on heating/cooling or field application/removal, which are separated by thermal/field hysteresis[22]. Thus, by minimizing thermal hysteresis a larger overlap is achieved, maximizing the RC in a compound.

In order to check if the phase compatibility enhancement under pressure is particular to the $Ni_2Mn_{1.4}In_{0.6}$ composition or if it is a general property of NiMn-based Heusler alloys presenting martensitic magnetostructural phase transitions, we measured magnetization under hydrostatic pressure for the $Ni_{1.8}Co_{0.2}Mn_{1.4}In_{0.6}$ alloy. Just like $Ni_2Mn_{1.4}In_{0.6}$, $Ni_{1.8}Co_{0.2}Mn_{1.4}In_{0.6}$ also shows a martensite phase transition from a FM cubic austenite to an AFM 3M monoclinic martensite phase around 200 K with approximately 28 K thermal hysteresis. The $\lambda_2$ for this alloy at atmospheric pressure was calculated from temperature dependent neutron diffraction data (see Supplementary Material) to be 0.9899, deviating by approximately 1% from unity, while $Ni_2Mn_{1.4}In_{0.6}$ ($\lambda_2 = 1.0070$) deviates by 0.7%. As can be seen in Fig. 4, the effect of hydrostatic pressure on the thermal hysteresis of $Ni_{1.8}Co_{0.2}Mn_{1.4}In_{0.6}$ is even more drastic. The phase transitions at cooling and heating are much more sensitive to pressure and shift to higher temperatures at a rate of 6.8 K/kbar and 8.4 K/kbar, respectively. Consequently, thermal hysteresis is decreased by half of the ambient pressure value upon application of 10 kbar (28.7 K for P = 0 and 14.3 K for P = 10 kbar), demonstrating that this behavior is more widely found in NiMn-based magnetocaloric Heusler alloys.

To conclude, we show from pressure dependent magnetization in $Ni_2Mn_{1.4}In_{0.6}$ and $Ni_{1.8}Co_{0.2}Mn_{1.4}In_{0.6}$ magnetocaloric Heusler alloys that the thermal hysteresis across the martensite transition is linearly decreased upon the application of hydrostatic pressure. The origin of this behavior is investigated using high pressure XRD which reveals that the lower latent heat and hysteresis minimisation with pressure are linked with the geometrical compatibility condition: with increasing pressure the system more closely satisfies the $\lambda_2 = 1$ condition. Thus the geometrical compatibility between martensite and austenite phases at the martensite phase transition in magnetocaloric Heusler alloys can be enhanced and tuned by physical pressure. This leads to a large reduction of the phase transformation hysteresis. Our present study underlines the importance of pressure as an external parameter to overcome the large hysteresis and energy barrier problem in phase transforming magnetic Heusler materials aiming at applications in magnetic refrigeration.

**Acknowledgments:** Authors would like to thank Dr. Emmanuelle Suard from the Institute Laue-Langevin, Grenoble-France, for support during neutron diffraction experiments. S.S. thanks Science and Engineering Research Board of India for the Ramanujan fellowship.

*luana.caron@cpfs.mpg.de

†sanjay.singh@cpfs.mpg.de

**Figures:**

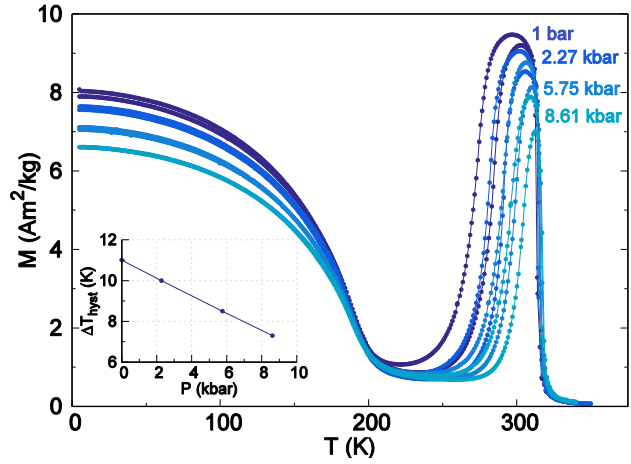

**Fig.1:** Field cooled heating and cooling magnetization measurements at different pressures and an applied magnetic field of 0.01 T for Ni$_2$Mn$_{1.4}$In$_{0.6}$. The inset shows the pressure dependence of the thermal hysteresis of the martensitic phase transition.

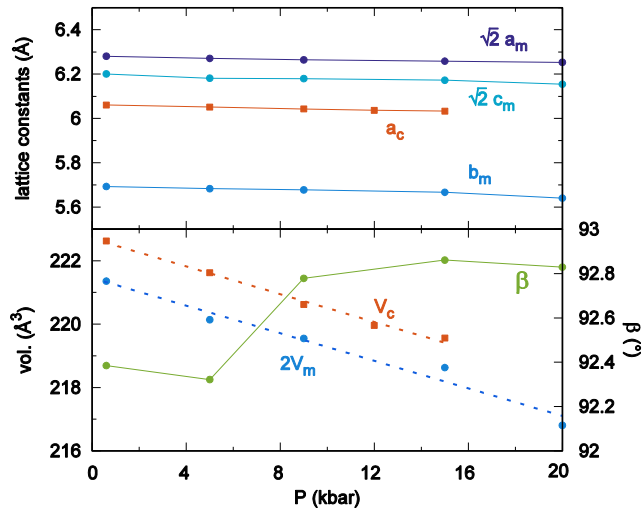

**Fig.2:** Lattice parameters and volume of the cubic austenite (a$_c$ and V$_c$ measured above the transition at 320 K) and monoclinic martensite (a$_m$, b$_m$, c$_m$, β and V$_m$ measured below the martensite transition at 240 K) phases under hydrostatic pressure for Ni$_2$Mn$_{1.4}$In$_{0.6}$. The dotted lines are linear fits of the volume data.

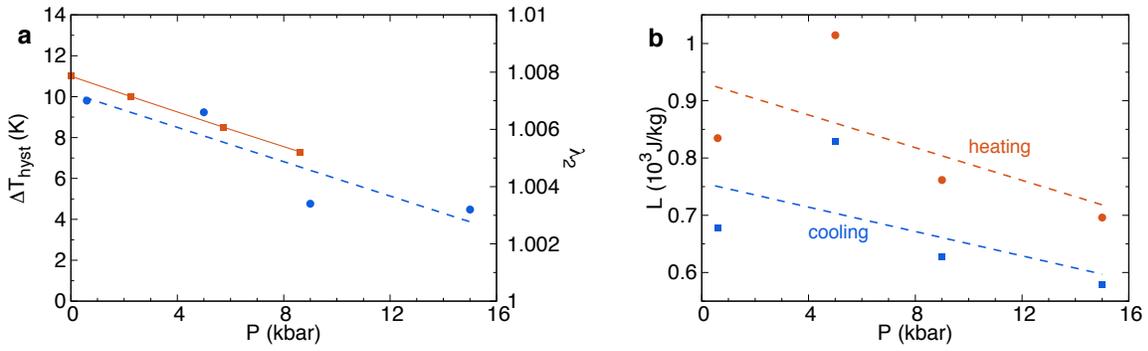

**Fig.3:** (**a**). Comparison of thermal hysteresis (orange squares) and the middle eigenvalue $\lambda_2$ (blue circles) as a function of pressure for $Ni_2Mn_{1.4}In_{0.6}$. The lines are linear fits of the presented data. (**b**). Latent heat calculated using the Clausius-Clapeyron relation for $Ni_2Mn_{1.4}In_{0.6}$ as a function of pressure. Since $dT/dP$ and $T_t$ are different on heating and cooling, two sets of latent heat values are obtained corresponding to the two transitions. The lines are linear fits of the presented data.

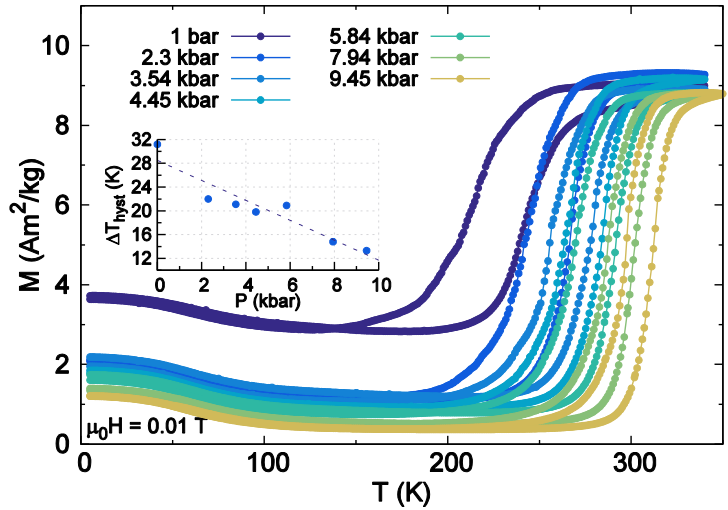

**Fig.4:** Field cooled heating and cooling magnetization measurements at different pressures and an applied magnetic field of 0.01T for $Ni_{1.8}Co_{0.2}Mn_{1.4}In_{0.6}$. The inset shows the pressure dependence of the thermal hysteresis of the martensitic phase transition.

# Supplementary Material

# Minimizing hysteresis in martensite phase transforming magnetocaloric Heusler alloys

Luana Caron,[1, a)] Parul Devi,[1] Alexandre M. G. Carvalho,[2] Claudia Felser,[1] and Sanjay Singh[1, 3, b)]

[1)] *Max Planck Institute for Chemical Physics of Solids, Nöthnitzer Str. 40, 01187 Dresden, Germany*

[2)] *Laboratório Nacional de Luz Síncrotron (LNLS), Centro Nacional de Pesquisa em Energia e Materiais (CNPEM), CEP 13083-970, Campinas, São Paulo, Brazil*

[3)] *School of Materials Science and Technology, Indian Institute of Technology (Banaras Hindu University), Varanasi-221005, India*

(Dated: 13 June 2018)

## I. SAMPLE PREPARATION AND CHARACTERIZATION

Polycrystalline samples of composition $Ni_2Mn_{1.4}In_{0.6}$, and $Ni_{1.8}Co_{0.2}Mn_{1.4}In_{0.6}$ were prepared from high purity elements by arc melting (repeated several times after flipping the button to ensure homogeneity) and subsequent annealing in a quartz ampoule under Ar atmosphere and quenched in an ice water mixture. The annealing temperature and time were 973 K for 72 h and 1173 K for 24 h for $Ni_2Mn_{1.4}In_{0.6}$ and $Ni_{1.8}Co_{0.2}Mn_{1.4}In_{0.6}$, respectively.

Magnetic measurements under hydrostatic pressure were performed in a home-made CuBe piston-cylinder type pressure cell built to fit the sample space of the MPMS XL magnetometer. A small polycrystalline piece (mass 2.75mg) was measured. Silicon oil is used as pressure transmitting medium. A small piece of Sn is loaded with the sample and functions as a manometer. Thus, the pressure inside the cell is inferred from the dependence of the superconducting transition of Sn, which occurs around 3.7 K at 1 bar.[1] The pressures reported for the magnetic measurements in this work are corrected for the pressure drop that occurs on cooling the pressure cell from room temperature to 3.7 K. The pressure drop is estimated from a separate calibration measurement to be around 2 kbar, obtained by measuring the $T_C$ of high purity MnAs for which the pressure dependence is well-known.[2]

Temperature dependent XRD under hydrostatic pressure measurements were performed at the XDS beamline of the Brazilian Synchrotron Light Laboratory. For this measurement the $Ni_2Mn_{1.4}In_{0.6}$ sample was ground into powder and annealed at 973 K for 10 h followed by quenching into water. From the annealed powder, particles under 10 $\mu$m in size were selected by sieving and loaded on a diamond anvil cell. Small ruby grains were loaded along with the sample so that the fluorescence lines could be used to determine the pressure in the sample space. The pressure transmitting medium used was a mixture of four parts methanol to one part ethanol. The pressure cell was loaded into a cryostat for temperature control while the pressure was changed in situ using a gas membrane system. The wavelength of the radiation used was 0.619921 Å. The data was acquired by a 2D detector and integrated using $LaB_6$ as a calibration standard in the software FIT2D.[3] The XRD patterns obtained were fitted using the Le Bail[4] algorithm as implemented in the Jana2006 software package.[5]

## II. MIDDLE EIGENVALUE CALCULATION

For a reversible transformation, the middle eigenvalue $\lambda_2$ of the cubic to monoclinic transformation matrix $U$ should approach unity. The transformation matrix with the axis of monoclinic symmetry along the $\langle 100 \rangle_{cubic}$ direction is given by:[6]

$$U = \begin{pmatrix} \tau & \sigma & 0 \\ \sigma & \rho & 0 \\ 0 & 0 & \delta \end{pmatrix}$$

Where the elements in the matrix are defined as:

$$\tau = \frac{\alpha^2 + \gamma^2 + 2\alpha\gamma(sin\beta - cos\beta)}{2\sqrt{\alpha^2 + \gamma^2 + 2\alpha\gamma sin\beta}}$$

$$\rho = \frac{\alpha^2 + \gamma^2 + 2\alpha\gamma(sin\beta + cos\beta)}{2\sqrt{\alpha^2 + \gamma^2 + 2\alpha\gamma sin\beta}}$$

$$\sigma = \frac{\alpha^2 + \gamma^2}{2\sqrt{\alpha^2 + \gamma^2 + 2\alpha\gamma sin\beta}}$$

And $\delta = {}^b/_{a_0}$, $\alpha = {}^{a\sqrt{2}}/_{a_0}$, $\gamma = {}^{c\sqrt{2}}/_{Na_0}$ are a function of the cubic lattice parameter $a_0$ and of the monoclinic lattice parameters $a$, $b$ and $c$ as well as the monoclinic angle $\beta$ and the degree of modulation $N$.

The lattice parameters and angle for the cubic and monoclinic structures were obtained from the patterns presented in Fig. 1 using the Le Bail pattern fitting method.[4] XRD patterns were taken well under the magneto-structural phase transition at 240 K and at the ferromagnetic austenite phase at 320 K on cooling mode. Notice that, at 20 kbar the cubic phase is no longer observed at 320 K and only the monoclinic phase is present.

————————
a)caron@cpfs.mpg.de
b)singh@cpfs.mpg.de

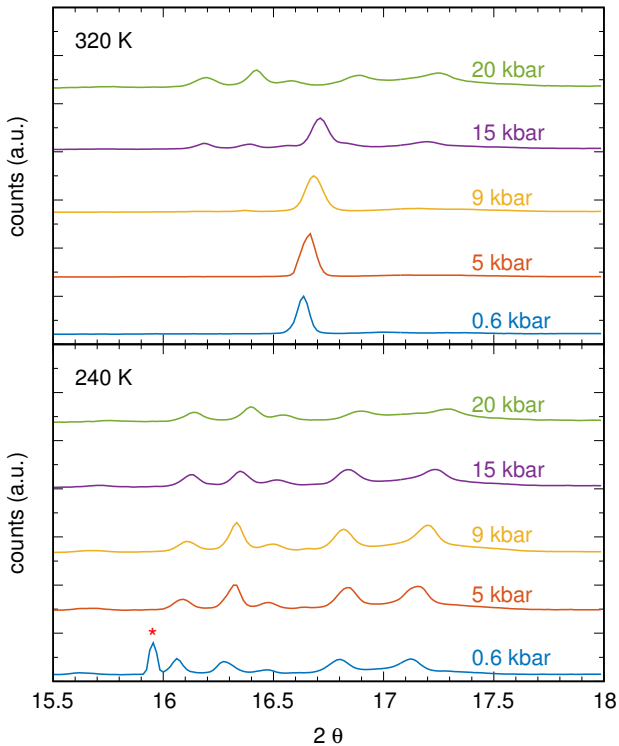

FIG. 1. X-rays diffraction under hydrostatic pressures up to 20 kbar at 320 K and 240 K for $Ni_2Mn_{1.4}In_{0.6}$. The red asterisk marks a spurious peak, probably due to the pressure cell gasket.

## III. NEUTRON DIFFRACTION ON $Ni_{1.8}Co_{0.2}Mn_{1.4}In_{0.6}$

Neutron diffraction measurements on $Ni_{1.8}Co_{0.2}Mn_{1.4}In_{0.6}$ were carried out in the austenite (300 K) and martensite (3 K) phases (see Fig. 2) at the D2B high-resolution neutron powder diffractometer (ILL, Grenoble). The powder sample was loaded in a vanadium cylindrical sample holder. The data were collected using a neutron wavelength of 1.59 Å in the high-intensity mode. The LeBail refinement of the powder diffraction patterns was performed using the JANA2006 software package.[5] The refined lattice parameters were 5.9893 Å at 300 K (cubic austenite phase) and a = 4.4022 Å , b = 5.5407 Å, c = 4.3216 Å and $\beta$ = 94.2410 at 3 K (monoclinic 7M modulated martensite phase). Using these lattice parameters the calculated value of $\lambda_2$ is 0.9899.

## IV. LATENT HEAT CALCULATION

The latent heat of the structural martensitic phase transition can be calculated from the experimental data using the Clausius Clapeyron relation:

$$\frac{\rho\,(\Delta V/V)}{\Delta S_t} = \left(\frac{\partial T_t}{\partial P}\right)_H$$

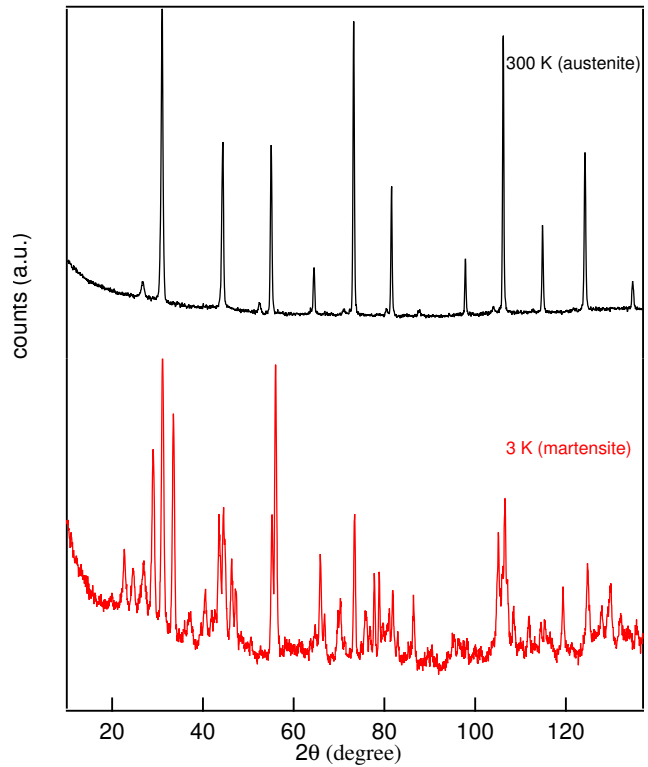

FIG. 2. Neutron diffraction measurements for $Ni_{1.8}Co_{0.2}Mn_{1.4}In_{0.6}$ in the austenite phase at 300 K and in the monoclinic phase at 3 K.

where $\Delta S_t$ is the entropy change due to the structural phase transition in the absence of field, $\Delta V/V$ is the relative volume change at the phase transition, $\left(\frac{\partial T_t}{\partial P}\right)_H$ is the shift of the phase transition with pressure at a given field and $\rho$ is the density of the material. For $Ni_2Mn_{1.4}In_{0.6}$ $\rho = 8.231.10^3 kg/m^3$.

Since $\Delta S_t = L/T_t$, where L is the latent heat and $T_t$ the transition temperature, the latent heat can be calculated from the high pressure crystallographic and magnetization data. Note that, the cooling and heating transitions shift at different rates with pressure, and thus have different $T_t$, reflecting a different energy barriers and thus L at the transition depending on the direction it is crossed.